\documentclass[osajnl,twocolumn,showpacs,superscriptaddress,10pt]{revtex4-1} %% use 11pt for Applied Optics
\usepackage{amsmath,amssymb,graphicx}

\usepackage{mathtools}
\usepackage{xfrac}
\begin{document}

\title {Gas-phase microresonator-based comb spectroscopy without an external pump laser}

\author{Mengjie Yu}
\email{my2473@columbia.edu}
\affiliation{Department of Applied Physics and Applied Mathematics, Columbia University, New York, NY 10027}
\affiliation{School of Electrical and Computer Engineering, Cornell University, Ithaca, NY 14853}

\author{Yoshitomo Okawachi}
\affiliation{Department of Applied Physics and Applied Mathematics, Columbia University, New York, NY 10027}

\author{Chaitanya Joshi}
\affiliation{Department of Applied Physics and Applied Mathematics, Columbia University, New York, NY 10027}
\affiliation{School of Applied and Engineering Physics, Cornell University, Ithaca, NY 14853}

\author{Xingchen Ji}
\affiliation{Department of Electrical Engineering, Columbia University, New York, NY 10027}
\affiliation{School of Electrical and Computer Engineering, Cornell University, Ithaca, NY 14853}

\author{Michal Lipson}
\affiliation{Department of Electrical Engineering, Columbia University, New York, NY 10027}

\author{Alexander L. Gaeta}
\affiliation{Department of Applied Physics and Applied Mathematics, Columbia University, New York, NY 10027}

\begin{abstract}We present a novel approach to realize microresonator-comb-based high resolution spectroscopy that combines a fiber-laser cavity with a microresonator. Although the spectral resolution of a chip-based comb source is typically limited by the free spectral range (FSR) of the microresonator, we overcome this limit by tuning the 200-GHz repetition-rate comb over one FSR via control of an integrated heater. Our dual-cavity scheme allows for self-starting comb generation without the need for conventional pump-cavity detuning while achieving a spectral resolution equal to the comb linewidth. We measure broadband molecular absorption spectra of acetylene by interleaving 800 spectra taken at 250-MHz per spectral step using a 60-GHz-coarse-resolution spectrometer and exploits advances of integrated heater which can locally and rapidly change the refractive index of a microresonator with low electrical consumption (0.9 GHz/mW), which is orders of magnitude lower than a fiber-based comb. This approach offers a path towards a simple, robust and low-power consumption CMOS-compatible platform capable of remote sensing. \end{abstract}

%\ocis{(320.6629) Supercontinuum generation; (190.4390) Integrated optics.}% REPLACE WITH CORRECT OCIS CODES FOR YOUR ARTICLE
                          % NOTE: \ocis{} IS ALIASED TO \pacs{} BUT MUST
                          % FORMAT THE TERMS CORRECTLY FOR EACH JOURNAL

\maketitle %% required

%\section{Introduction}

The past decade has witnessed the emergence of chip-scale optical frequency comb (OFC) sources based on a high-$\textit{Q}$ microresonator pumped with a continuous-wave (CW) laser. Compared to conventional OFC technology based on modelocked lasers, microresonators offer full wafer-scale integration and enable octave-spanning OFC's at low optical power level through parametric frequency conversion \cite{Brasch, Okawachi} and dispersion engineering that enables operation over different spectral windows \cite{Okawachi_shaping}. Such minature OFC's have been demonstrated in a wide range of microresonator platforms from the visible \cite{Lee}, near-infrared (near-IR) \cite{Herr, Brasch, Okawachi_shaping, Del'Haye, Jung, Jung2, Okawachi, Saha, Joshi, Yi, Hausmann,Savch,Xue,Volet}, and mid-infrared (mid-IR) \cite{Luke, Wang, Yu, Griffith}. Recently, applications based on microresonators have been unlocked beyond fundamental laboratory study, including microwave generation \cite{Liang}, dual-comb spectroscopy \cite{Yu2,Suh_dual,Pavlov,Dutt}, light detection and ranging \cite{Trocha,Suh_ranging}, optical frequency synthesizer \cite{Spencer}, terabit coherent communications \cite{Koos}, and astrocomb for exo-planet searches \cite{Suh_astro, Obrzud}. 

A major application of chip-scale OFC's is as a broadband optical source for massively parallel measurements of diverse molecules (Fig. 1). While fiber-based combs are a mature technology, the fine comb line spacing (100 MHz) requires a high-resolution spectrometer to directly resolve individual lines. In contrast, microresonator-base OFC's have a much larger spacing (from a few GHz to 1 THz) which enables comb-resolved detection using only a coarse-resolution spectrometer. This offers a significantly faster acquisition rate and potential for combining with an on-chip spectrometer \cite{Kita}. While comb spacings of $\sim$ 10 GHz have been demonstrated \cite{Suh_dual, Yi}, most microresonator platforms have inherently large comb line spacings which makes them unsuitable for most gas-phase spectroscopy that requires MHz-level resolution \cite{Newbury}. In addition, precise control of both the pump laser frequency and the cavity resonance is needed for comb generation, modelocking, and stabilization \cite{Herr, Del'Haye, Yu_scanning}. 

In this paper, we address both challenges using a fiber-microresonator dual-cavity scheme and a spectrometer with a coarse spectral resolution of 60 GHz. In our dual-cavity scanning comb (DCSC), the CW pump laser is replaced by a gain medium in a cavity that contains the Si$_{3}$N$_{4}$ microresonator \cite{Johnson, Pecc, Morand}. We scan the entire comb spectrum over one free spectral range (FSR, 195 GHz) by simply tuning the microresonator cavity resonance via an integrated heater, which removes the complexity of synchronization of pump laser tuning and cavity tuning \cite{Yu_scanning, Weiner_thermal}. Interleaving all the spectra \cite{Nathalie, Rutko, Diddams, Maslo} as the microresonator is tuned improves the spectral resolution to 250 MHz. We show that this system can be used for relatively high-spectral-resolution absorption measurements of acetylene gas at 400 Torr and 6 Torr. Such DCSC offers the potential for a turn-key, on-chip broadband spectrometer with fast acquisition speed suitable for trace gas sensing. 

\begin{figure}
\centering
\centerline{\includegraphics[width=8cm]{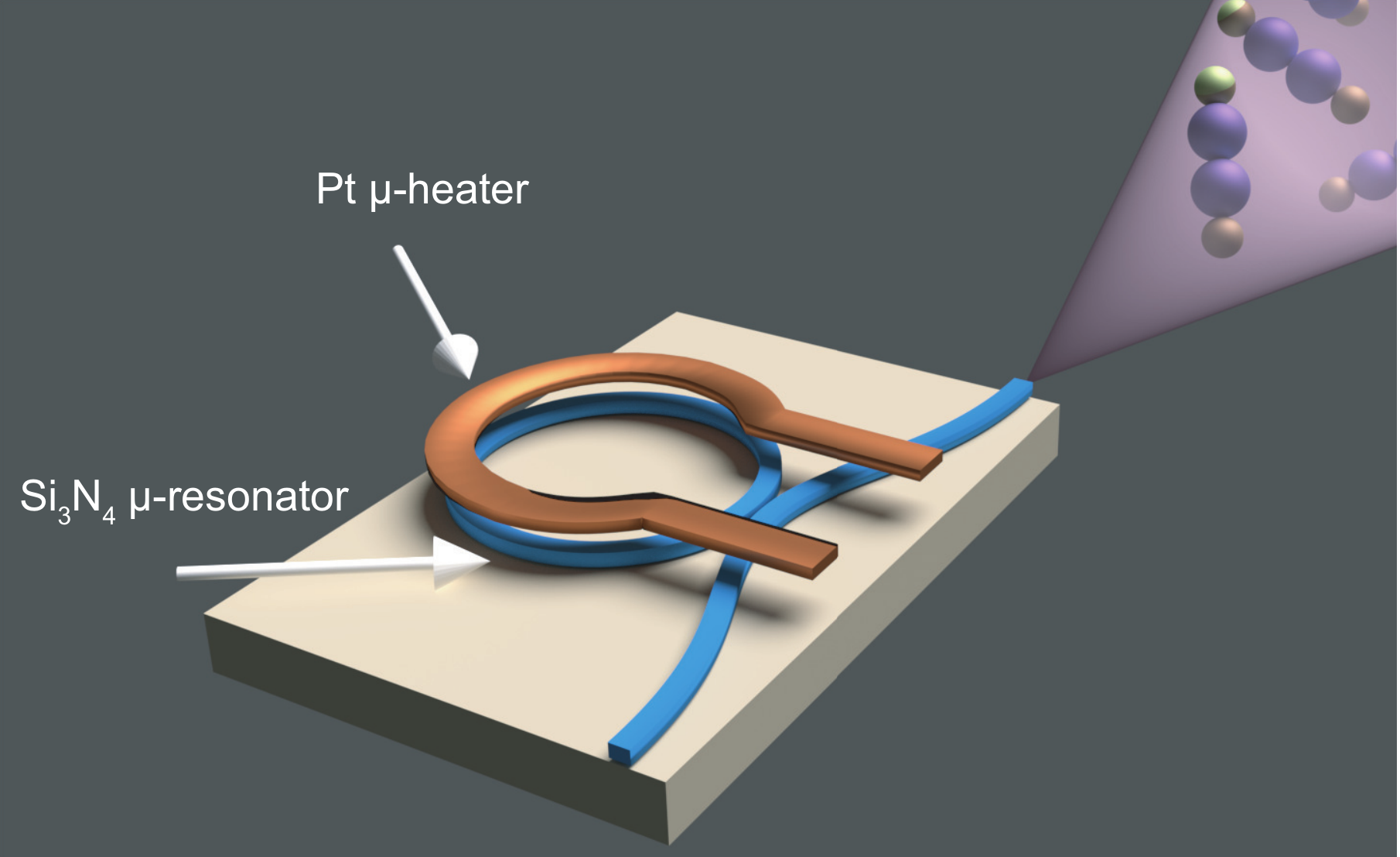}}
\caption{Microresonator-based molecular spectroscopy. A Si$_{3}$N$_{4}$ microresonator (with bus waveguide) is drawn with an integrated heater on top. The waveguides are cladded with silicon dioxide of 2.5 $\mu$m thickness (not shown), above which the integrated heater is fabricated. 
}
\label{Fig1}
\end{figure}

Our experimental setup is shown in Fig. 2. We use an oxide-cladded Si$_{3}$N$_{4}$ microresonator with a radius of 120 $\mu$m and a loaded $\textit{Q}$ of 1 million as the Kerr comb platform. The microresonator is engineered to have anomalous group-velocity-dispersion for the fundamental transverse electric modes with a cross section of 730 $\times $ 2200 nm. An integrated platinum resistive heater \cite{Joshi, Weiner_thermal} is fabricated on top of the oxide cladding to locally change the temperature near the microresonator with a resistance of 260 $\Omega$ (Fig. 1). The bus waveguide forms part of the external cavity and couples to the Si$_{3}$N$_{4}$ microresonator, similar to the scheme in Ref. 34. The external cavity also includes an erbium-doped fiber amplifier (EDFA), polarization components, and an optical isolator. The generated optical spectrum is measured with an optical spectrum analyzer (OSA) after a 90/10 coupler. By setting the polarization to quasi-TE and increasing the EDFA gain, an OFC spectrum spanning 1450 - 1700 nm is achieved [Fig. 3(a)] at an EDFA power of 200 mW. The optical power in the bus waveguide is 40 mW due to a 7-dB coupling loss into the chip. Due to the large Purcell factor of the microresonator, the EDFA preferentially amplifies the oscillating modes that lie within a microresonator resonance, and the resulting lasing mode serves as the pump for the parametric oscillation in the microresonator \cite{Johnson}. In contrast to the conventional CW pump laser, comb generation is self-starting once the EDFA reaches the threshold power, without the need to sweep the laser into resonance \cite{Herr}. Moreover, it does not suffer from the disruption of comb generation due to the drifts of the relative pump-cavity detuning. Most importantly, this dual-cavity configuration is ideal for tuning the comb spectrum since only the position of the microresonator resonance needs to be tuned and the "pump" mode follows accordingly.

\begin{figure}
\centering
\centerline{\includegraphics[width=8cm]{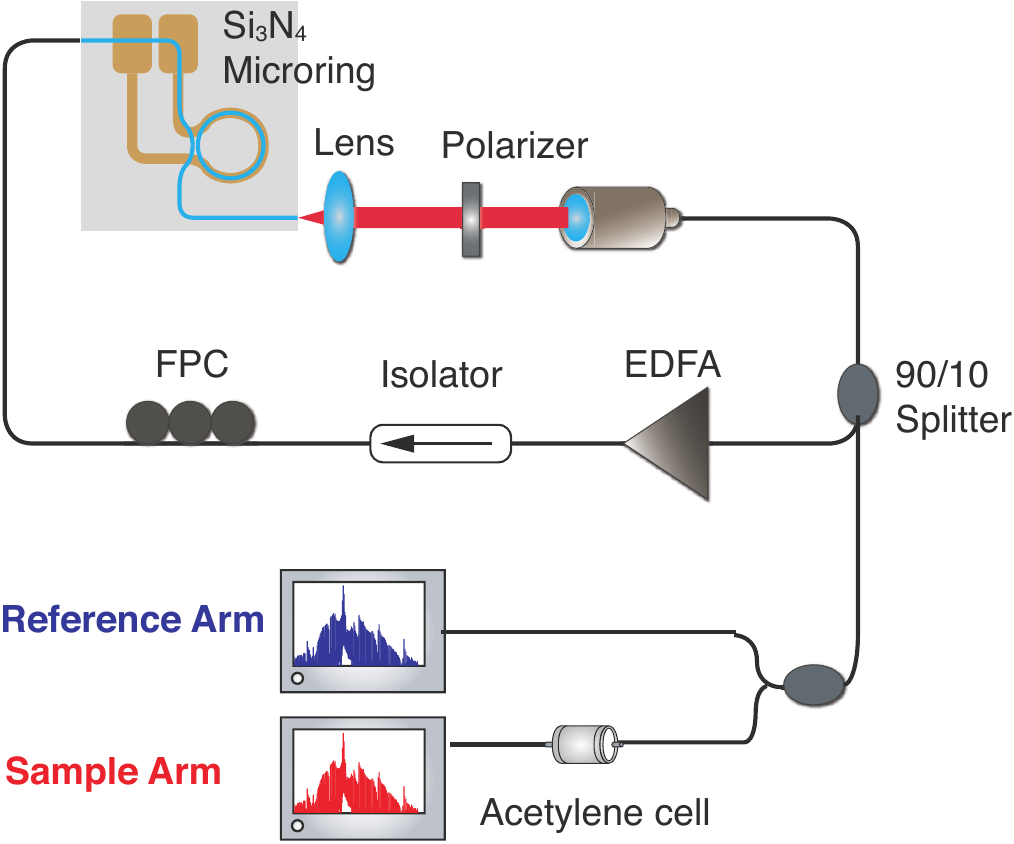}}
\caption{Experimental setup for dual-cavity scanning comb. The first cavity (in gray region) is the Si$_{3}$N$_{4}$ microresonator for optical parametric oscillation. The second cavity is the external loop formed by the bus waveguide, polarization components, EDFA and an optical isolator. 10$\%$ of the power in the external cavity is coupled out via a 90/10 coupler. We use two optical spectrum analyzers, one for measuring the acetylene absorption spectrum and the other for calibration. FPC: fiber polarization controller, EDFA: erbium-doped fiber amplifier. 
}
\label{Fig2}
\end{figure}

In our experiment, we control the voltage (power) of the integrated heater to shift the cavity resonance via the thermal-optic effect once we generate the comb. The relationship between the heater power and the tuning frequency is characterized by a high resolution OSA (1.25 GHz) and is used for programming the scanning process. The frequency accuracy is also limited by the calibration of the initial spectrum. We tune the entire comb spectrum over one full FSR (195 GHz) via tuning the integrated heater, and the process is fully driven by a computer algorithm. The heater voltage is initially set at 3V to offset the cavity temperature. Figure 3(a) shows 11 different optical spectra over the scanning range at an OSA resolution of 25 GHz. The zoom-in spectra [Fig. 3(b)] shows a smooth power flatness of one comb line over the entire scanning range. The appearance of several spectral dips in Fig. 3(a) result from the localized dispersion perturbation due to mode crossings \cite{Sven}, which causes low signal to noise ratio (SNR) of the comb lines nearby. In the radio-frequency spectrum of the output [Fig. 3(c)], the beat notes separated by 5.2 MHz correspond to the external fiber cavity length of 39.5 meters and the 250-MHz bandwidth corresponds to the microresonator linewidth. This also indicates the OFC is not operating in a modelocked state and has a spectral coherence up to the resonance linewidth of the microresonator (250 MHz), which is the fundamental limit of the spectral resolution of our system for spectroscopy. However, the resolution could be improved to several MHz level using a Si$_{3}$N$_{4}$ microresonator with a much higher $\textit{Q}$ recently reported in Refs. 43 and 44. Using a low gain EDFA to reduce the fiber cavity length could also improve the coherence of the OFC \cite{Pecc}. Lastly, the heater efficiency is characterized to be 0.9 GHz/mW, as shown in Fig. 1(e), corresponding to about 2.2 mW/$^{\circ}$C. Thus, we require only 220 mW of electrical power to shift the comb by 195 GHz (one full FSR), which corresponds to a temperature change of 100 $^{\circ}$C. The power consumption is orders of magnitude lower than that of a fiber-based comb. 

\begin{figure}
\centering
\centerline{\includegraphics[width=8cm]{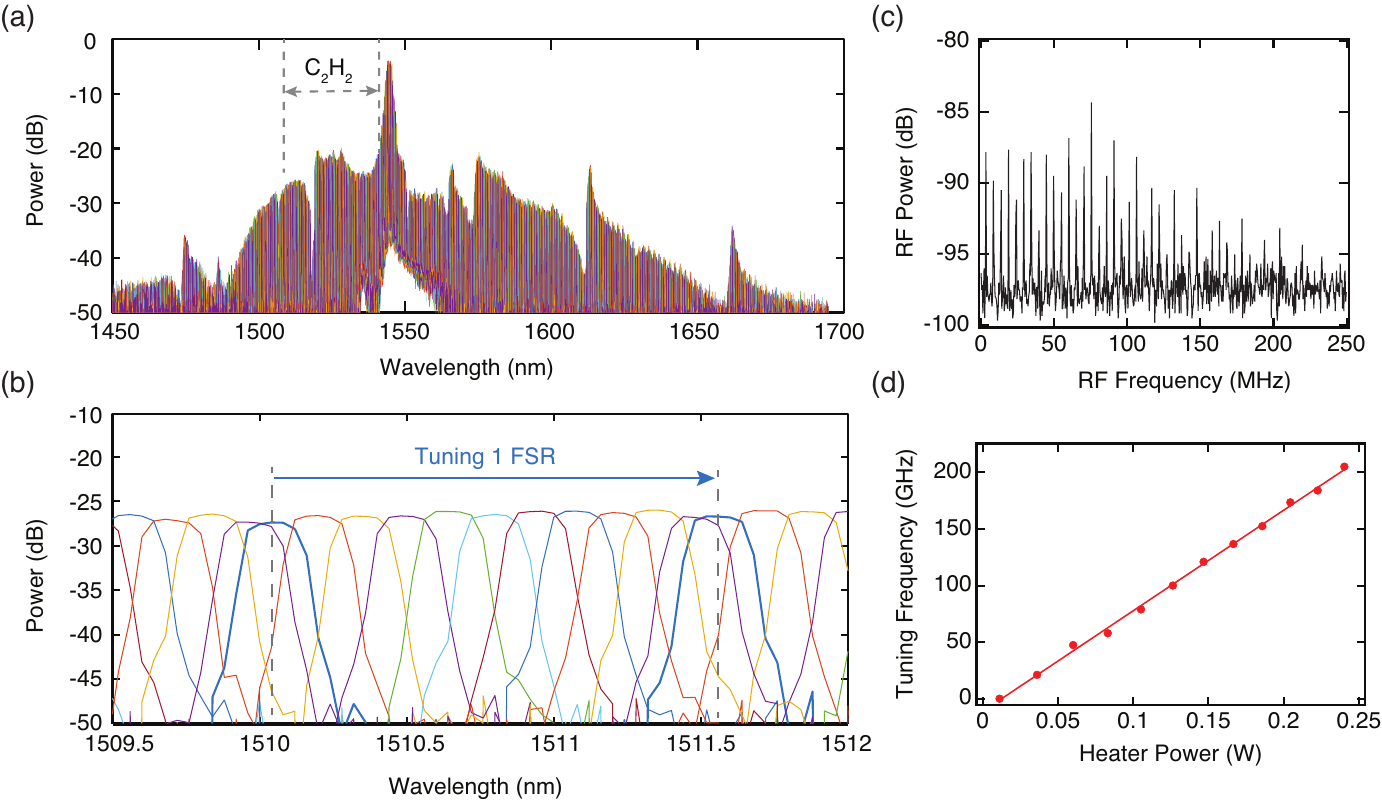}}
\caption{Characterization of the comb tuning. (a) The recorded frequency comb spectra at different heater voltages. The spectrum spans from 1450 to 1700 nm at an EDFA output power of 200 mW. Multiple mode crossings cause the spectral dips and affect the bandwidth. The absorption of acetylene resides between the grey dashed lines which spans about 40 nm. (b) Zoom-in spectra of (a), which shows the comb line is tuned over one full free spectral range of 195 GHz. (c) The radio-frequency spectrum of the comb. The line spacing is 5.2 MHz, which corresponds to an external cavity length of 39.5 meters. The detector bandwidth is 10 GHz. (d) The tuning frequency as a function of heater power consumption. The efficiency is 0.9 GHz/mW. 
}
\label{Fig3}
\end{figure}

As a proof-of-principle, we utilize the DCSC for measuring the absorption spectrum of the P and R branches of gas-phase acetylene in the $\upsilon_{1} + \upsilon_{3}$ band, which spans over part of the comb spectrum (1508 - 1543 nm) in the near-IR. As shown in Fig. 2, we split the output into the reference and sample arms, each of which is sent to an OSA with a low resolution of 60 GHz. The reference arm is used for calibration of the spectrum. We use a 5.5-cm-long acetylene cell at 400 Torr in the sample arm. The scanning process is programmed to be 250 MHz per step. We acquire 800 spectra with a speed of 0.5 s per acquisition, which is limited by the slow readout time from the OSA. The absorption is calculated based on the spectral measurement in the reference and sample arms. Figure 4(a) shows the measured absorption spectrum of acetylene [black curve], which is recorded by interleaving 800 spectra taken with 250-MHz steps. The repetition rate change with the temperature is taken into account to improve the frequency accuracy \cite{Yu_scanning}. We compare the measured spectrum to that [inverted red curve] based on the HiTran database, which is in good agreement. A total of 22 (labelled) out of 150 comb lines is used for the absorption measurement in the P and R branches, each of which is shaded in different colors in the Fig. 4(a). The small grey area between 7th comb line and 8th comb line is due to a missing comb line, which is attributed to a mode crossing effect, therefore the 7th comb line is tuned slightly more to cover its area. Figure 4(b) shows the measured transmission of the 8th and 19th comb lines that are each tuned over one FSR range. Several absorption features with different depth and width are clearly captured. Figure 4(c) shows one measurement of the R(9e) line of the $\upsilon_{1} + \upsilon_{3}$ band as compared to HiTran. A standard deviation of the residual is 1.4 $\times$ 10$^{-2}$, which is largely limited by the SNR of the corresponding comb line. A drop port device for outputting the OFC spectrum should improve the coupling efficiency and the SNR's. 

\begin{figure}
\centering
\centerline{\includegraphics[width=8cm]{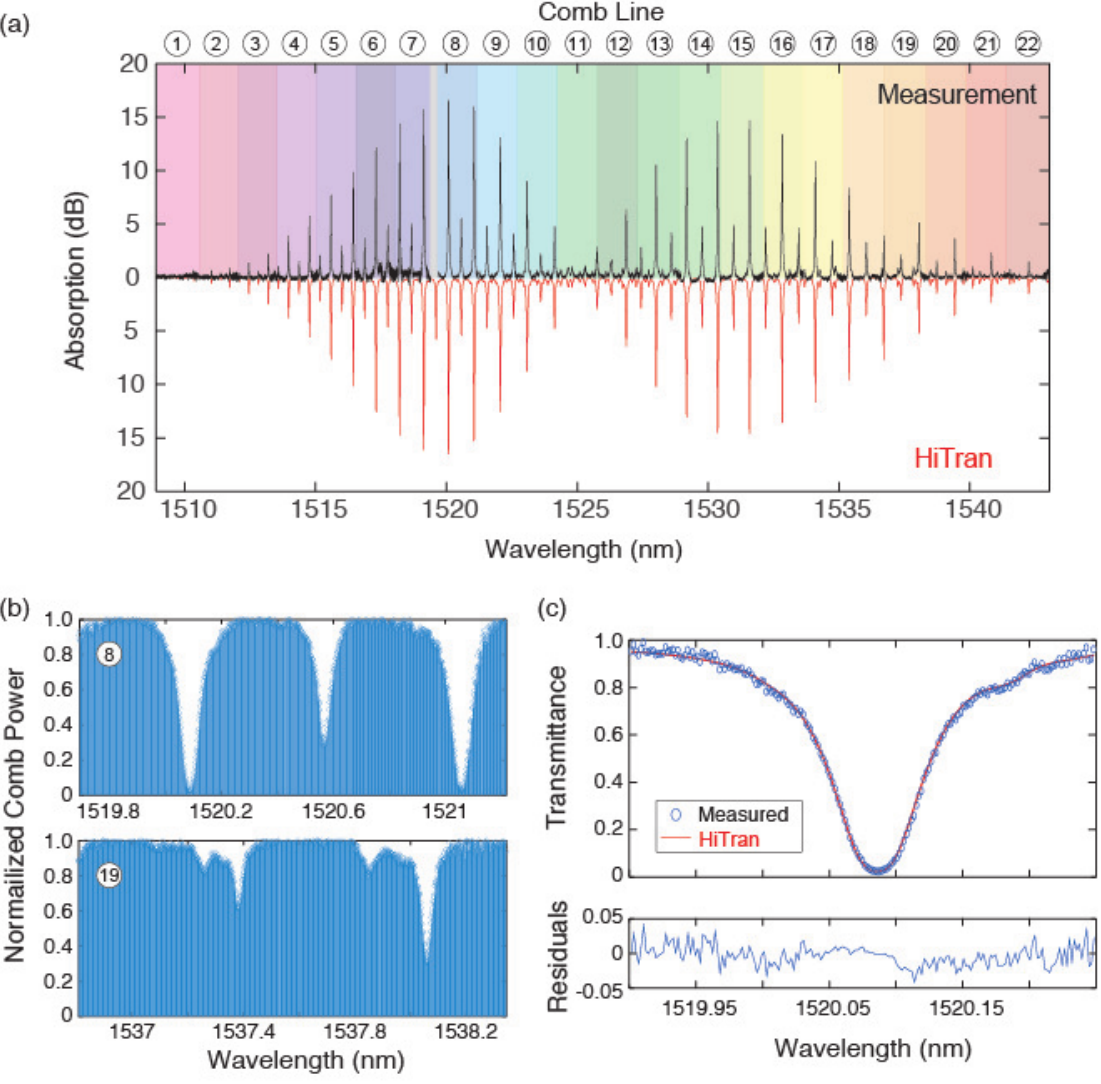}}
\caption{Spectral measurements of acetylene. The gas cell is 400-Torr pure acetylene with a length of 5.5 cm. (a) The absorption spectrum is calculated by interleaving 800 spectra at 250 MHz [black curve]. 22 comb lines of the entire optical spectrum is overlapped with molecular absorption. The 1-FSR tuning range of each comb line are shaded in different colors. The grey area (no data points) is due to a missing comb line. The HiTran data is calculated using the gas cell condition and plotted for comparsion [inverted red curve]. (b) The comb line transmission measured by the 8th and 19th comb lines according to (a). (c) The measured transmittance of the R(9e) line of the $\upsilon_{1} + \upsilon_{3}$ band as compared to HiTran. The standard deviation of the residuals is 1.4 $\times$ 10$^{-2}$.
}
\label{Fig4}
\end{figure}

We also apply this technique for measuring an acetylene cell at a lower pressure of 6 Torr with an effective length of 11 cm. Figure 5(a) shows the measured absorption spectrum [red curve], and we compare this to a model spectrum [inverted black curve] based on a convolution of the microresonator line-shape and the absorption spectrum from HiTran database, and good agreement is observed. Due to the fact that the comb linewidth is comparable to the absorption linewidth at this pressure, the measured absorption feature becomes weaker which causes some weak absorption peaks to be smeared as seen from Fig. 5(a), which could be addressed by using a higher $\textit{Q}$-factor microresonator. In addition, the scanning step can be easily tuned in order to optimize for different gas-cell conditions. For example, Fig. 5(b) shows the measurement of the P(7e) line of the $\upsilon_{1} + \upsilon_{3}$ band at different scanning steps of 10, 20, 50, 100 and 250 MHz. We measure a 380-MHz half-width-half-maximum width of the absorption line which also agrees well with the HiTran data. 

\begin{figure}
\centering
\centerline{\includegraphics[width=8cm]{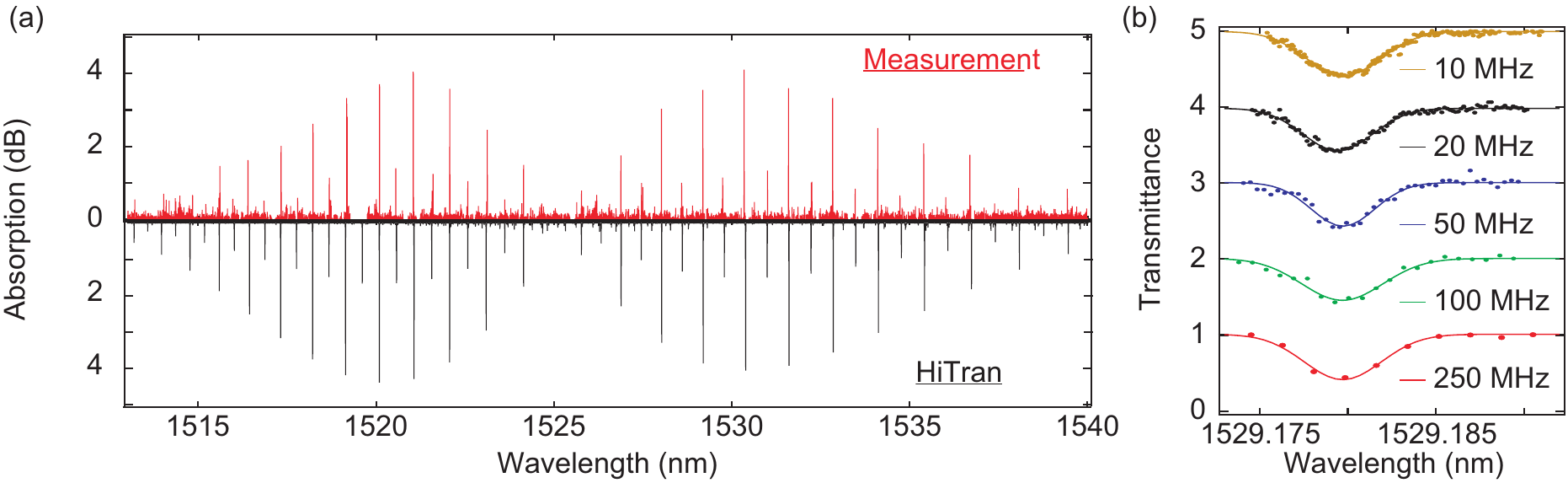}}
\caption{Measurement of a low-pressure cell of acetylene at 6 Torr. The effective cell length is 11 cm. (a) Absorption spectrum by interleaving 800 spectra at a 250-MHz step, similar to Fig. 4(a). (b) Transmittance of the P(7e) line measured at different scanning steps of 10, 20, 50, 100 and 250 MHz. 
}
\label{Fig5}
\end{figure}

Due to a fast response time of integrated heaters (about 10 $\mu$s), the scan can be performed significantly faster than the piezo-tuning speed of CW lasers ($>$ 1 ms). While the acquisition speed is currently limited by the slow readout time of the OSA, by using a low-resolution grating along with an InGaAs camera ($<$ 100 $\mu$s per image), the data-acquisition rate can be significantly increased to enable real-time output. As a proof-of-principle, we filtered out the comb line near an absorption line using a 1-nm bandpass filter and sent it to a photodetector (Thorlabs PDA10CS, bandwidth 17 MHz). We sweep the frequency by 5 GHz by sending a periodic triangular voltage function to the heater at rates of 1 Hz, 10 Hz, 100 Hz and 200 Hz. The recorded single-shot oscilloscope traces are shown in Fig. 6 with time normalized to one period, corresponding to a tuning speed of 10 GHz/s, 100 GHz/s, 1 THz/s and 2 THz/s.  At 1 THz/s, the absorption features start to show distortion, which indicates that our scheme can potentially achieve a total acquisition time of $<$ 2 s to complete the current 200-GHz tuning if combined with a grating and a line camera. While microresonator-based dual-comb spectroscopy \cite{Dutt, Suh_dual, Yu2} can achieve much faster acquisition speed (2 $\mu$s \cite{Yu2}), it requires delicate control of two modelocked frequency combs with external pump lasers, and its spectral resolution suffers from the large optical line spacing (10 - 200 GHz) which is more suitable for condensed phase studies. Our DCSC approach offers a much simpler and more robust approach for gas-phase spectroscopy with a reasonably fast acquisition rates. 

\begin{figure}
\centering
\centerline{\includegraphics[width=8cm]{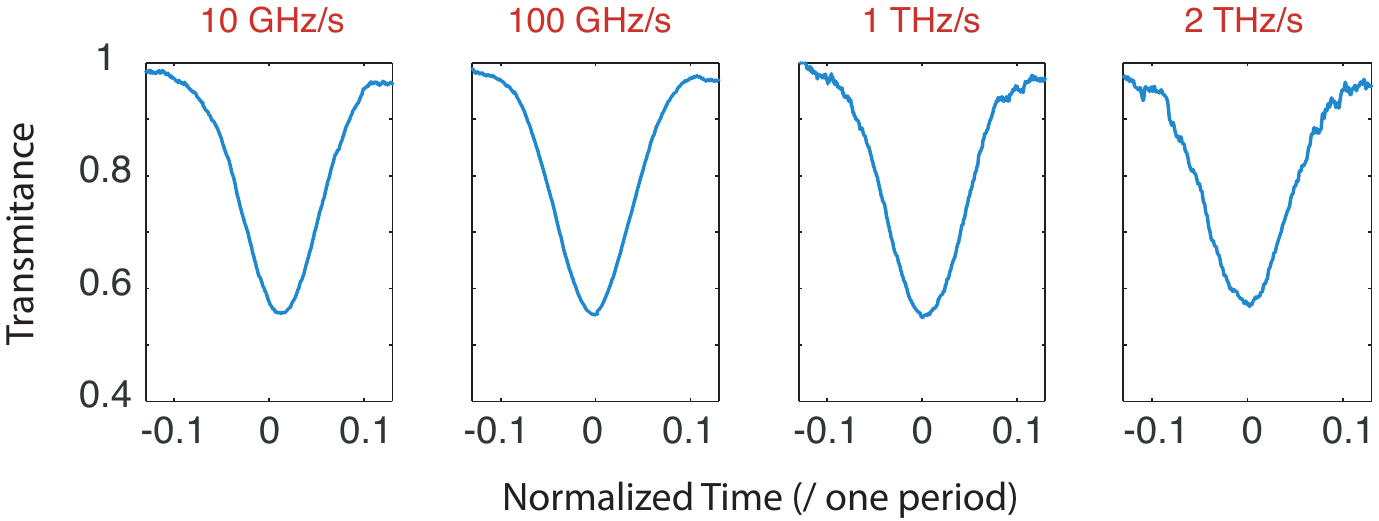}}
\caption{Measurement of one absorption line at different scanning speeds (single-shot). The corresponding comb line is filtered out using a 1-nm bandpass filter after the acetylene cell and then sent to a fast photodetector. The heater is programmed to sweep a range of 5 GHz at 200 Hz, 100 Hz, 10 Hz, and 1 Hz (from top to bottom), which corresponds to a tuning speed of 2 THz/s, 1 THz/s, 100 GHz/s and 10 GHz/s. The time axis of recording time traces is normalized with respective to the scanning period. 
}
\label{Fig6}
\end{figure}

In summary, we demonstrate a new approach to gas-phase spectroscopy using microresonator-based OFC technology. We program an integrated heater to tune the comb spectrum over one full FSR in a dual-cavity scheme and measure the molecular fingerprint of acetylene in the $\upsilon_{1} + \upsilon_{3}$ band. Parallel detection of multiple trace gases can benefit from the broadband comb generated from microresonators since only 15$\%$ of the spectrum was used here for the acetylene measurements. Integrated heaters offer a high tuning efficiency at low electrical power consumption, a rapid tuning speed, and the capability of full integration with microresonators. Moreover, our DCSC can be extended to the important mid-IR regime, where molecules have fundamental vibrational transitions, by replacing EDFA with a quantum cascade amplifier. Combining with the silicon microresonator shown in Ref. 18, a quantum or interband cascade amplifier with a minimum power of 35 mW (at 2.8 $\mu$m) or 80 mW (at 3 $\mu$m) could potentially be applied for gas sensing in the mid-IR using our configuration. In conclusion, with a higher $\textit{Q}$-factor microresonator \cite{Xuan, Ji}, a chip-scale gain medium \cite{Stern, Stern2} and an on-chip spectrometer \cite{Kita}, we envision a lab-on-chip system with real-time output and high spectral resolution suitable for gas sensing over a wavelength range far larger than what could be achieved with a tunable laser. 

\textbf{Funding.} Defense Advanced Research Projects Agency (DARPA) (W31P4Q-15-1-0015), Air Force Office of Scientific Research (AFOSR) (FA9550-15-1-0303) and National Science Foundation (NSF) (ECS-0335765). 
\\
\\
\textbf{Acknowledgment.} The research was funded by Defense Advanced Research Projects Agency (DARPA) (W31P4Q-15-1-0015), Air Force Office of Scientific Research (AFOSR) (FA9550-15-1-0303) and National Science Foundation (NSF) (ECS-0335765). This work was performed in part at the Cornell Nano-Scale Facility, a member of the National Nanotechnology Infrastructure Network, which is supported by the NSF.

% Bibliography

\clearpage
\newpage
%\pagebreak

\end{document}